\documentclass[12pt,preprint]{aastex}

\shorttitle{Cosmic-ray acceleration at ultrarelativistic shocks}
\shortauthors{J. Niemiec \& M. Ostrowski}
\usepackage{natbib}

\begin{document}

\title{COSMIC-RAY ACCELERATION AT ULTRARELATIVISTIC SHOCK WAVES: \\
 EFFECTS OF A ``REALISTIC'' MAGNETIC FIELD STRUCTURE}

\author{Jacek Niemiec}
\affil{Department of Physics and Astronomy, Iowa State University, Ames,
IA 50011, USA}
\affil{and Institute of Nuclear Physics PAN, ul. Radzikowskiego 152,
 31-342 Krak\'{o}w, Poland}
\email{niemiec@iastate.edu}
\and
\author{Micha\l{} Ostrowski}
\affil{Astronomical Observatory, Jagiellonian University, \\
ul. Orla 171, 30-244 Krak\'{o}w, Poland}

\begin{abstract}
First-order Fermi acceleration processes at ultrarelativistic ($\gamma \sim 5 - 30$) 
shock waves are studied with the method of Monte Carlo simulations. The accelerated 
particle spectra are derived by integrating the exact particle trajectories 
in a turbulent magnetic field near the shock. The magnetic field model upstream 
of the shock assumes finite-amplitude perturbations within a wide wavevector 
range and with a predefined wave power spectrum, imposed on the mean field 
component inclined at some angle to the shock normal. The downstream field 
structure is obtained as the compressed upstream field. We show that the main 
acceleration process at superluminal  shocks is the particle compression 
at the shock. Formation of energetic spectral tails is possible in a limited 
energy range only for highly perturbed magnetic fields. Cutoffs in the spectra 
occur at low energies within the resonance energy range considered. These 
spectral features result from the anisotropic character of particle transport 
in the magnetic field downstream of the shock, where field compression 
produces effectively 2D perturbations. We also present results for parallel 
shocks. Because of the turbulent field compression at the shock, the acceleration 
process becomes inefficient for larger turbulence amplitudes, and features 
observed in oblique shocks are recovered in this case. For small-amplitude 
perturbations, particle spectra are formed in the wide energy range, and 
modifications of the acceleration process due to the existence of long-wave 
perturbations are observed, as reported previously for mildly relativistic 
shocks. The critical turbulence amplitude required for 
efficient acceleration at parallel shocks decreases with increasing shock 
Lorentz factor $\gamma$. In both subluminal and superluminal shocks, an 
increase of $\gamma$ leads to steeper spectra with lower cut-off energies. 
The spectra obtained for the ``realistic'' background conditions assumed in our 
simulations do not converge to the ``universal'' spectral index claimed in the 
literature. Thus the role of the first-order Fermi acceleration in 
astrophysical sources hosting relativistic shocks requires serious reanalysis.
\end{abstract}


\section{INTRODUCTION}
At relativistic shock waves the bulk flow velocities are comparable to particle
velocities. This leads to highly anisotropic particle distributions at the shock.
Unlike the case of nonrelativistic shocks, acceleration processes are 
very sensitive to the background conditions, in particular to the structure of 
the perturbed (``turbulent'') magnetic field and the details of particle-wave 
interactions. These are, however, poorly known and modeling of the particle 
acceleration must rely on simplifying assumptions regarding particle scattering 
and transport. Here we consider the first-order Fermi acceleration processes, 
involving only the high-energy particles with gyroradii much greater than the 
thickness of the shock. 

The first consistent semianalytic method to study the acceleration process at 
parallel relativistic shock was proposed by \citet{kir87a} and it was further 
extended to treat more general conditions at the shock by \citet{hea88}. 
The method uses the stationary Fokker-Planck pitch-angle diffusion equation 
describing anisotropic particle distributions near the shock. 
Matching the upstream and downstream solutions of the diffusion equation at the 
shock yields both the power-law index of the resulting spectrum 
and the form of anisotropic particle angular distribution. The spectral indices 
for the phase space distribution function, $\alpha$, derived with this method 
for parallel relativistic shocks are different from the value of 
$\alpha_{\rm NR} = 4$ (or the equivalent {\it energy} spectral index 
$\sigma_{\rm NR} =\alpha_{\rm NR}-2= 2$) obtained 
for strong nonrelativistic shocks (e.g., Drury 1983) and depend on the wave 
power spectrum of the magnetic field perturbations ($\equiv$ the form of the
pitch-angle diffusion coefficient). The same approach, augmented by the 
assumption of magnetic moment conservation for particles interacting with the 
shock, was used to study acceleration processes in oblique subluminal shocks by 
\citet{kir89}. They proved that oblique shocks provide harder spectra than 
the parallel ones and the spectral index $\alpha$ approaches $3$ when the 
shock velocity along magnetic field lines is close to the speed of light. 
In the weakly perturbed conditions they assumed, the particle density upstream of 
the shock can be much larger than its downstream value \citep{ost91}, which 
results from the very effective multiple reflections of anisotropically 
distributed upstream particles from the compressed downstream magnetic field.

The validity of the semianalytic approximation is restricted to the case of a 
weakly perturbed magnetic field, where cross-field diffusion does not play a 
significant role. As discussed by \citet{beg90}, application of such conditions 
in perpendicular (superluminal) shocks does not allow for the diffusive 
acceleration process, since particles are not able to cross the shock front 
repeatedly. Instead, the shock-drift acceleration process, involving a single 
transmission of upstream cosmic ray particles through the shock 
(Webb, Axford, \& Terasawa 1983; Drury 1983) shapes the particle spectrum in
superluminal shocks. As a result, the distribution of accelerated particles 
downstream of the shock is the upstream particle distribution compressed 
and shifted to higher energies. Because of particle angular distribution anisotropy,  
the mean energy gain of particles transmitted through the shock is larger than 
the one expected from the adiabatic compression.  
 
Efficient particle cross-field scattering can be accomplished if finite-amplitude
MHD waves occur near the shock. They may preexist in the turbulent ambient medium
upstream of the shock, but they can be also generated by the accelerated 
particles themselves \citep[as in nonrelativistic shocks, see, e.g.,][]{dru83,luc00}. 
In such conditions, investigations of the acceleration processes must rely on 
numerical methods. 
A role of finite-amplitude magnetic field perturbations in forming particle 
spectra at relativistic shocks was investigated by a number of authors using 
Monte Carlo particle simulations 
\citep[e.g.,][]{ell90, ost91, ost93, bal92, nai95, bed96, bed98}. 
A direct dependence of the (power-law) particle spectra on the 
assumed conditions near the shock was proved. For different mean magnetic field 
inclinations
with respect to the shock normal and different amplitudes of perturbations, 
the spectra can be either steep or flat, and variations of the particle spectral
index can be non-monotonic at the transition from weakly perturbed conditions 
to highly nonlinear magnetic field perturbations \citep{ost91,ost93}. For highly
turbulent magnetic fields, power-law spectra can be also formed at 
superluminal shocks. They are, however, very steep for mildly perturbed 
conditions.
 
The power spectra of realistic perturbed magnetic fields can differ by such 
parameters as the amplitude and the spectral index for the considered 
power-law distributions, but also the wavevector range of the distribution and
its possible anisotropy or nonuniformity. The first step to study such 
realistic features was presented 
for mildly relativistic shocks in a recent paper \citep{nie04},%
\footnote{The two earlier papers by \citet{bal92} and \citet{ost93}, which 
incorporated some ``realistic'' features of the magnetic field near the shock,
had to apply a very simplified modeling of the turbulent field structure 
(see \citet{nie04}; here we want to correct some inaccuracies concerning the 
simulations of \citet{bal92}, which appeared in our paper: rather than having 
unrelated upstream and downstream turbulent fields, their magnetic fields were
consistently transformed from upstream to downstream frame at the shock boundary;
secondly, the same turbulent field was used throughout the integration of  
particle trajectories --- there was no selection of a new turbulent field at each
particle shock crossing).} where we 
considered the power-law wave spectra upstream of the shock within a relatively 
wide range of wavevectors, with only static perturbations involved. 
The perturbations are imposed on the mean field component, inclined at a 
given angle to the shock normal. The downstream magnetic field 
structure was derived by assuming a simple shock compression of the upstream 
field, thus preserving the continuity of the perturbed magnetic field 
across the shock. By choosing different amplitudes and spectral 
indices of the turbulence and different mean field inclinations, we 
were able to study the acceleration processes in a variety of background 
conditions, including both sub- and superluminal shocks with varying amount of 
background perturbations. The model allowed us also to investigate the  
role --- in the shock acceleration --- of short, resonance and long waves at 
different particle energies, to better understand the microphysics of the 
acceleration processes. The study showed that the particle spectra generally 
diverge from a simple power law; the exact shape of the spectrum depends on both 
the amplitude and the wave power spectrum of the magnetic field perturbations. 
We reported and discussed such features as spectrum hardening before the cutoff 
at high particle energies in oblique subluminal shocks, and the steepening of the
spectrum and the cutoff formation in the resonance energy range in superluminal 
shocks. In the latter case, the role of the long-wave magnetic field 
perturbations proved to be critical for the generation of the power-law spectral 
tails. For parallel shocks, we showed that the presence of 
finite-amplitude magnetic field perturbations leads to the formation of locally 
oblique field configurations at the shock and the respective magnetic field 
compressions. This implies a qualitative modification of the particle acceleration 
process by introducing some features present in oblique shocks. As a result, 
nonmonotonic variations of the particle spectral index with the turbulence 
amplitude were observed. 

In the present paper we slightly modify the above approach to study shocks with 
larger Lorentz factors $\gamma \gg 1$. Besides several detailed results, 
we report significant difficulties to form the wide-energy range power-law 
particle spectral tails 
in quasi-perpendicular shocks.%
\footnote{One should note that nearly all magnetic field 
configurations in ultrarelativistic shocks lead to the perpendicular 
(superluminal) shock structure.}
In most cases, including highly perturbed conditions at the shock, an energy 
cutoff appears in the particle spectrum within the considered resonance energy 
range for particle-wave interactions. 
Moreover, even for parallel shocks with the studied magnetic 
field structures, we do not observe spectral index convergence to the 
``universal'' value discovered by Bednarz \& Ostrowski (1998) and 
Gallant \&  Achterberg (1999) \citep[see also discussion by][]{ostb02}.
These results are important in a wide context of interpreting the observational
data on astrophysical sources involving highly relativistic plasma flows/shocks,
such as active galactic nuclei, micro-quasars, gamma-ray bursts, as well as for 
the ultra--high-energy cosmic ray production problem.  
In a forthcoming paper we plan to extend the present analysis to 
situations including the generation of short-wave turbulence at the shock.
Preliminary results of this work are presented in \citet{nie05}.

In what follows, $c$ is the speed of light. All calculations are
performed in the respective local plasma (upstream or downstream) rest frames. 
The upstream (downstream) quantities are labeled with the index `1' (`2'). 
We consider ultrarelativistic particles with $p=E$. In units we use in our 
simulations, a particle of an unit energy moving in an uniform mean upstream 
magnetic field $B_{0,1}$, has the unit maximum (for $p_{\perp}=E$) gyroradius 
$r_g(E=1)=1$ and the respective resonance wavevector $k_{res}(E=1) = 2\pi$. 

\section{SIMULATIONS}

The Monte Carlo approach we use here has been described in detail in 
\citet{nie04}. Below we sketch the most important features of the method and 
discuss the modifications introduced in the present paper.

We consider a planar relativistic shock wave propagating with velocity $u_1$ 
(Lorentz factor $\gamma_1$) with
respect to the magnetized upstream plasma. The upstream magnetic field is 
assumed to consist of the uniform component {\bf\em B}$_{0,1}$, inclined at 
some angle $\psi_1$ to the shock normal, and finite-amplitude perturbations 
imposed upon it. The irregular component has either a flat $(F(k)\sim k^{-1})$ 
or a Kolmogorov $(F(k)\sim k^{-5/3})$ wave power spectrum in a wide 
wavevector range $(k_{min}, k_{max})$. The downstream field structure, together 
with the downstream flow velocity $u_2$, are obtained from hydrodynamic shock 
jump conditions derived for the cold electron-proton plasma upstream of the 
shock. We use approximate formulae derived by Heavens \& Drury (1988) to 
calculate the shock rest frame compression ratio $R = u_1/u_2$ 
($R=3$ for $\gamma_1\rightarrow\infty$). The acceleration
process is studied by following exact particle trajectories in the perturbed 
magnetic field near the shock, without a simplifying hybrid approach used by 
Niemiec \& Ostrowski (2004). Because the irregular field component is assumed 
to be static in the upstream and downstream plasma rest frames, the second-order 
Fermi acceleration is excluded from our considerations. 
Radiative (and other) energy losses are also neglected in this modeling.

\subsection{Perturbed Magnetic Field Structure}

The ``turbulent'' ($\equiv$ perturbed) magnetic field component upstream of the 
shock is modeled as a superposition of static sinusoidal waves of finite 
amplitude. In the magnetic field related {\it primed} coordinate 
system,\footnote{In the upstream plasma rest frame the $x$-axis of the (unprimed) 
Cartesian coordinate system $(x, y, z)$ is perpendicular to the shock surface. 
The shock wave moves in the negative $x$-direction and the regular magnetic 
field {\bf\em B}$_{0,1}$
lies in the $x-y$ plane. The primed system $(x'$, $y'$, $z')$ is obtained from 
the unprimed upstream one by its rotation about the $z$-axis by an angle 
$\psi_1$, so that the $x'$-axis is directed along {\bf\em B}$_{0,1}$.}
they take the form:
\begin{equation}
\delta B_{x'} = \sum_{l=1}^{294} \delta B_{x'l} \,\sin (k_{x'y'}^{l} y' + 
k_{x'z'}^{l} z'), 
\end{equation}
where $(k_{x'y'}^{l})^{2} + (k_{x'z'}^{l})^{2}  = (k_{x'}^{l})^2$ (see below),
and analogously for $\delta B_{y'}$ and $\delta B_{z'}$ components. Such a form 
of $\delta${\bf\em B} for 3D turbulence \citep[see][]{gia94,mich97} ensures 
that $\nabla\cdot${\bf\em B}$=0$. The index $l$ 
enumerates the logarithmic wavevector range from which the wavevectors 
$k_{i}^{l} \: (i = x',y',z')$ are randomly drawn. The components 
$k_{x'y'}^{l}$, $k_{x'z'}^{l}$ of the wavevectors $k_{x'}^{l}$ are selected by 
choosing a random phase angle 
$\phi _{x'}^{l}$, so that $k_{x'y'}^{l}=k_{x'}^{l}\cos\phi_{x'}^{l}$ and 
$k_{x'z'}^{l}=k_{x'}^{l}\sin\phi_{x'}^{l}$, and analogously for other components.
The wavevectors span the range $(k_{min}, k_{max})$, where $k_{min} = 0.0001$ and
$k_{max} = 10$, and particle energies are formally related to these values by 
the resonance condition $k_{res} = 2\pi / r_g(E)$.
 
The respective wave amplitudes $\delta B_{il}$ are selected at random, subject 
to the constraint ${\delta B_{l}}^2 = \sum_{i} {\delta B_{il}}^2$, where the 
amplitudes $\delta B_{l}$ are chosen to reproduce the assumed turbulence power 
spectrum. In the wavevector range considered, it can be written as
\begin{equation} 
\delta B_{l}(k) = \delta B_{l}(k_{min}) \left( \frac{k}{k_{min}}
\right)^{(1-q)/2},
\end{equation}
where $q$ is the spectral index: $q=1$ corresponds to the flat spectrum and 
$q=5/3$ to the Kolmogorov one. The constant $\delta B_{l}(k_{min})$ is scaled 
to match the model parameter $\delta B / B_{0,1}$ defined upstream of the shock, 
where $\delta B \equiv [\sum_{l} {\delta B_{l}}^2]^{1/2}$ (or using the 
turbulence power spectrum $F(k)$, $\delta B^2=\int F(k)dk$). 

The downstream magnetic field structure is obtained as the compressed upstream 
field. According to the jump conditions at the shock \citep[e.g.,][]{hea88,kir99},
only field components perpendicular to the shock normal are compressed, and the 
relativistic compression factor $r=R\gamma_1/\gamma_2$ 
(where $\gamma_i=1/\sqrt{1-u_i^2} \; (i = 1,2)$) gives the compression ratio 
between the two plasma rest frames. One should note that in our approach the 
magnetic field lines are continuous across the shock. This allows one to study
upstream--downstream
correlations in particle motion introduced by the field structure, for different
levels of turbulence, and to study the influence of this factor in forming particle
spectra.

Although the Kolmogorov wave power spectrum considered in our modeling seems to 
properly describe magnetic field turbulence in a number of real astrophysical 
situations \citep[see, e.g., the discussion in][]{nie04}, the actual field 
structure near the shock front can be far more complex. Nonlinear interactions 
between accelerated particles and the background plasma initiate the generation of
MHD turbulence by plasma instabilities. Additionally, wave damping processes 
as well as nonlinear wave interactions influence the final turbulence spectrum. 
No currently used numerical approach allows one to study these processes in full, 
with involved nonlinearities and particle acceleration from thermal energies up 
to high-energy cosmic rays. Therefore, we limit our considerations to the 
test particle approach in the assumed background conditions. It restricts the 
validity of the study to the I-order Fermi acceleration process and particles 
with energies much higher than the ``thermal'' energies of the downstream plasma.

In the present simulations we assume that the shock front instabilities do not 
produce any additional fluctuations of the magnetic field. However, theoretical 
considerations \citep{med99} and numerical studies \citep[e.g.,][]{sil03,nis03,fre04} 
suggest that additional strong, small-scale field components are generated 
downstream of the shock due to the relativistic filamentation instability. 
The effects of these small-scale perturbations will be investigated in our 
forthcoming paper.

\subsection{Simulation Method and Particle Injection}

The particle equations of motion are integrated in the local (upstream or 
downstream) plasma rest frame, where the electric field vanishes. The simulation 
run is divided into cycles. At the beginning of each cycle, $N$ particles 
(usually $N=100$) are located at the shock front and Lorentz-transformed into the 
upstream plasma rest frame. The trajectory of each particle is then followed 
until it crosses the shock surface, at which point the Lorentz 
transformation to the downstream plasma rest frame is performed. At this side of 
the shock a free escape boundary is located ``far downstream'' from the shock at 
$x_{max}(E) = X_{max}\, \bar{r}_{g,2}(E_2)$ behind it (note that
$\bar{r}_{g,2}(E_2)$ is the particle 
gyroradius in the average downstream magnetic field). The selected value of the 
coefficient $X_{max}\in (5,100)$ depends on the chosen simulation parameters, and 
it is specified using numerical tests separately for each shock configuration.
This ensures that the results are not influenced by $X_{max}$. 
In the conditions considered by us, very few particles are able to travel/diffuse
 from such distance back to the shock. A particle trajectory downstream of
the shock is integrated until it crosses this escape boundary or reaches the 
shock front. After performing this procedure for all $N$ particles, a simulation 
cycle is finished. Then the trajectory splitting procedure is applied, which 
replaces all escaped particles with the ones still active in the acceleration 
process, with the respective partition of particle statistical weights. In this way 
particle spectra are derived with approximately the same accuracy in a wide energy 
range \citep[for details see][]{nie04}. The analogous subsequent cycles are 
repeated until either more than 90\% of 
particles escape through the downstream boundary in an individual simulation 
cycle, or all particles reach the assumed upper energy limit $E_{max,2}$ (given 
in the downstream rest frame), or the lower limit for the particle weight $w$ is
reached. The lower limit for $w$ is 
usually set to $10^{-6}$ of the initial weight $w_0$. The final spectra 
and angular distributions are averaged over several ($10$ - $50$) simulation
runs, each with statistically different sets of $N$ particles and different 
realizations of the perturbed magnetic field.

At the beginning of a simulation run the particles are injected at the shock 
front with uniform angular distribution within the range of 
$\cos\theta_2\in (-1,-u_2)$ in the downstream plasma rest frame ($\theta$ is the 
angle between the particle momentum and the shock normal), so that their momentum
vectors in the upstream rest frame are in the cone around the shock normal with 
the opening angle $\theta_c$ ($\sin\theta_c=1/\gamma_1$).
They are monoenergetic in the downstream rest frame, and the 
injection energy $E_i=E_{0,2}=0.1$ is such that particle resonance wavevectors 
$k_{res}(E_{0,2}) \gg k_{max}$. Such initial conditions correspond roughly to 
the particle injection from the thermal plasma. For ultrarelativistic shocks
considered here, the majority of the injected particles escape far 
downstream in the first simulation cycle. In order to improve the statistical 
accuracy of the spectra, we continue injecting particles in the first cycle until
$N$ of those which --- after transmission downstream --- succeed in reaching the 
shock front again are selected. These particles are used in a given simulation 
run and have the same initial weights $w_0$.

\subsection{Derivation of Particle Trajectories}

In the previous paper we developed a hybrid approach for modeling the motion of 
high-energy particles in the turbulent field including short MHD waves. It 
involves the exact integration of particle orbits in the perturbed field 
including long and resonance waves only, and the short-wave ($\lambda_s\ll r_g$)
influence on the trajectory is reproduced with a small-amplitude pitch-angle 
scattering term. A need for such approach arose due to excessive 
computation time required to derive exact trajectories in the magnetic field with
short-wave perturbations. This limitation becomes less severe for 
ultrarelativistic shock acceleration studies because for the high-$\gamma$ shocks
both upstream and downstream residence times are shorter than in the case of the
previously investigated mildly relativistic shocks, and for a particle of a given
energy these times decrease with $\gamma$. Thus, in the present paper we 
{\em do not} use the hybrid approach, and calculate particle orbits directly by 
integrating their equations of motion in the considered perturbed magnetic field 
near the shock.

In the computations we use the Runge-Kutta 5-th order method with the adaptive 
step size control (routine RKQS in: Press et al. 1992) for  accurate derivation 
of particle trajectories in non-uniform magnetic field with the optimized 
simulation time. The accuracy imposed in our simulations (controlled by the 
truncation error set to be less than 10$^{-7}$) ensures that the integration 
time step is much smaller than the residence time upstream and downstream of the 
shock.\footnote{This requirement limits the use of the hybrid method for 
ultrarelativistic shocks. The trajectory perturbations due to the short waves 
with $\lambda_s\ll r_g$ introduce diffusive modifications of particle trajectories 
on time scales much longer than $\tau_s \equiv \lambda_s/c$. When the residence 
time upstream or downstream of a high-$\gamma$ shock becomes comparable to 
$\tau_s$ the hybrid method involving a diffusive term cannot be used.} 
We checked by numerical tests, that the computed particle orbits are stable over 
more than the mean upstream and downstream residence time and the derived 
trajectories are reversible in time over their entire path. This proves that the 
numerical noise is maintained at the very low level, which is independently 
demonstrated by the accompanied diffusion coefficients derivations (see Appendix).

\section{RESULTS}

In the present work we use Monte Carlo simulations to study the role of the
realistic magnetic field features on the first-order Fermi acceleration processes
at ultrarelativistic shocks in the test particle approach. As pointed out by 
\citet{beg90}, most (essentially all) magnetic field configurations in the case 
of ultrarelativistic shocks lead to superluminal conditions for the acceleration 
process. However, our study includes the case of 
highly perturbed magnetic fields. Results of the modeling for superluminal shocks
are presented in \S 3.1. The structure of the compressed downstream magnetic 
field renders particle acceleration inefficient at such shocks --- even in 
conditions of highly perturbed magnetic fields the particle spectra do not extend
to very high energies, exhibiting cutoffs within the resonant range for particle-wave 
interactions. Parallel high-$\gamma$ shock configurations are studied in \S 3.2. 
We find that for larger amplitudes of the magnetic field perturbations, the 
spectral features similar to those observed in oblique shocks occur because of the 
effects of the magnetic field compression downstream of the shock. Only in the 
case of weakly perturbed fields, the flat power-law parts appear in the spectra
due to the presence of long-wave perturbations. 

Accelerated particle spectra formed at oblique superluminal shocks 
are presented in Figures \ref{wobl1} and \ref{wobl2}, and the spectra derived 
for parallel shock waves in Figures \ref{wpar1} and \ref{wpar2}. The background 
conditions for the respective spectra are indicated in the figures.
The spectra have been calculated for different upstream magnetic field 
perturbation amplitudes $\delta B/ B_{0,1}$. Some of the energetic particle
spectral tails have a power-law form for the conditions considered, and in 
these cases linear fits to the power-law parts of the spectra are presented 
with the values of the (phase-space) spectral index $\alpha$ given in the 
figures. Particles in the energy range indicated by arrows above the energy 
axis can interact with the magnetic field inhomogeneities with the resonant 
condition $k_{res}\equiv 2\pi/r_g$ satisfied, where 
$k_{min} \la k_{res} \la k_{max}$. These limits 
are not precise, they are calculated for the upstream magnetic field strength 
$B_{0,1}=1$ and particle transverse momenta  $p_{\perp}=p$. Note that the 
particle energies considered in the simulations can be approximately scaled to 
various conditions met in astrophysical shocks by the respective 
identifications of $B_{0,1}$, $\psi_1$, $\delta B$, $k_{min}$ and $k_{max}$.

\subsection{Superluminal Shock Waves}

\begin{figure}
\plotone{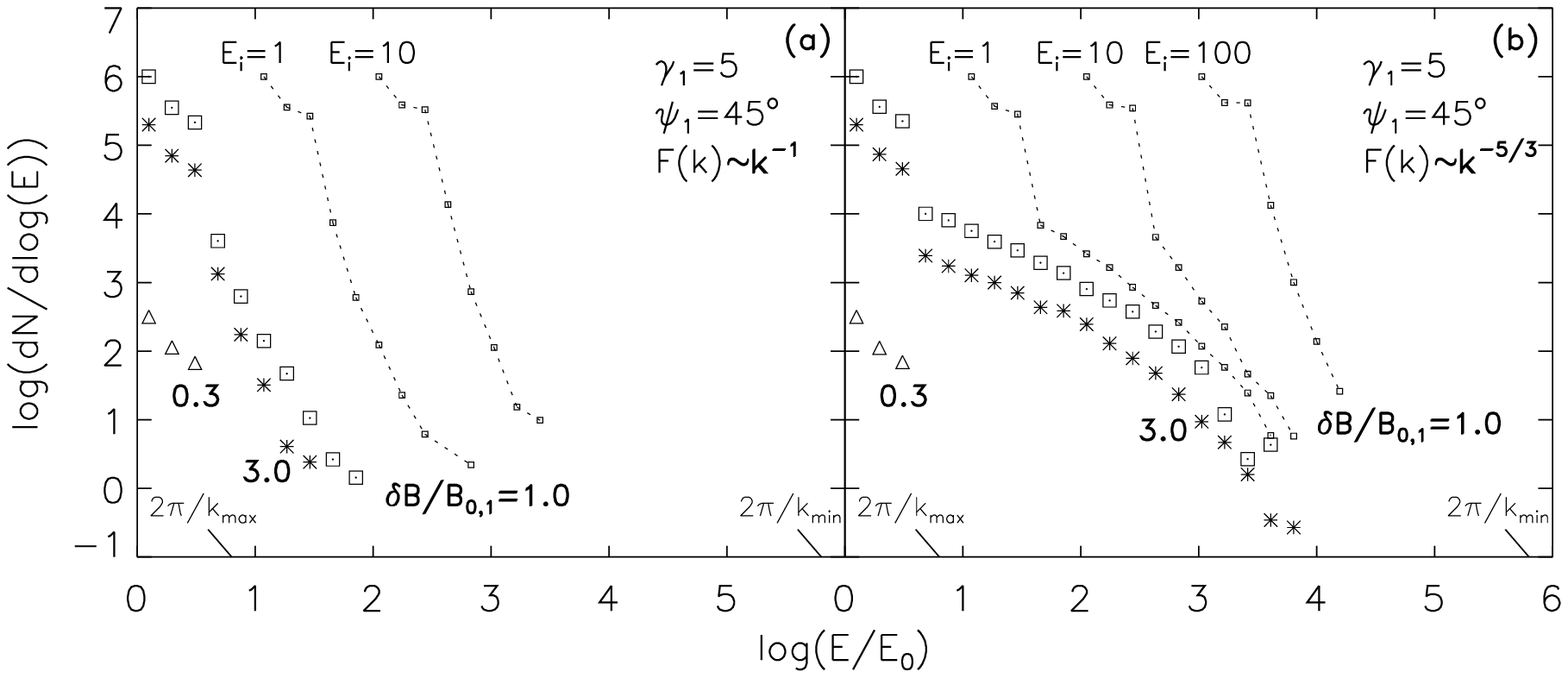}
\caption{Accelerated particle spectra at oblique superluminal shock waves with 
$\gamma_1=5$ and $\psi_1=45^o$ for (a) the flat and (b) the Kolmogorov wave 
spectrum of magnetic field perturbations. Any individual point in the spectrum 
represents a particle number ($\equiv$ weight) $dN$ recorded per 
a logarithmic energy bin. The upstream perturbation amplitude 
$\delta B / B_{0,1}$ is given near the respective results. The additional 
spectra obtained for higher injection energies $E_i = 1$, 10 
(and $100$ in Fig. \ref{wobl1}{\it b} ) are calculated 
for $\delta B / B_{0,1}=1.0$. Some spectra are vertically shifted for clarity. 
Particles in the range indicated as ($2\pi/k_{max}$, $2\pi/k_{min}$) can satisfy 
the resonance condition $k_{res} = 2 \pi / r_g(E)$ for some of the waves in the 
turbulence spectrum. The particle spectra are measured in the shock normal rest 
frame.
\label{wobl1}}
\end{figure}

\begin{figure}
\plotone{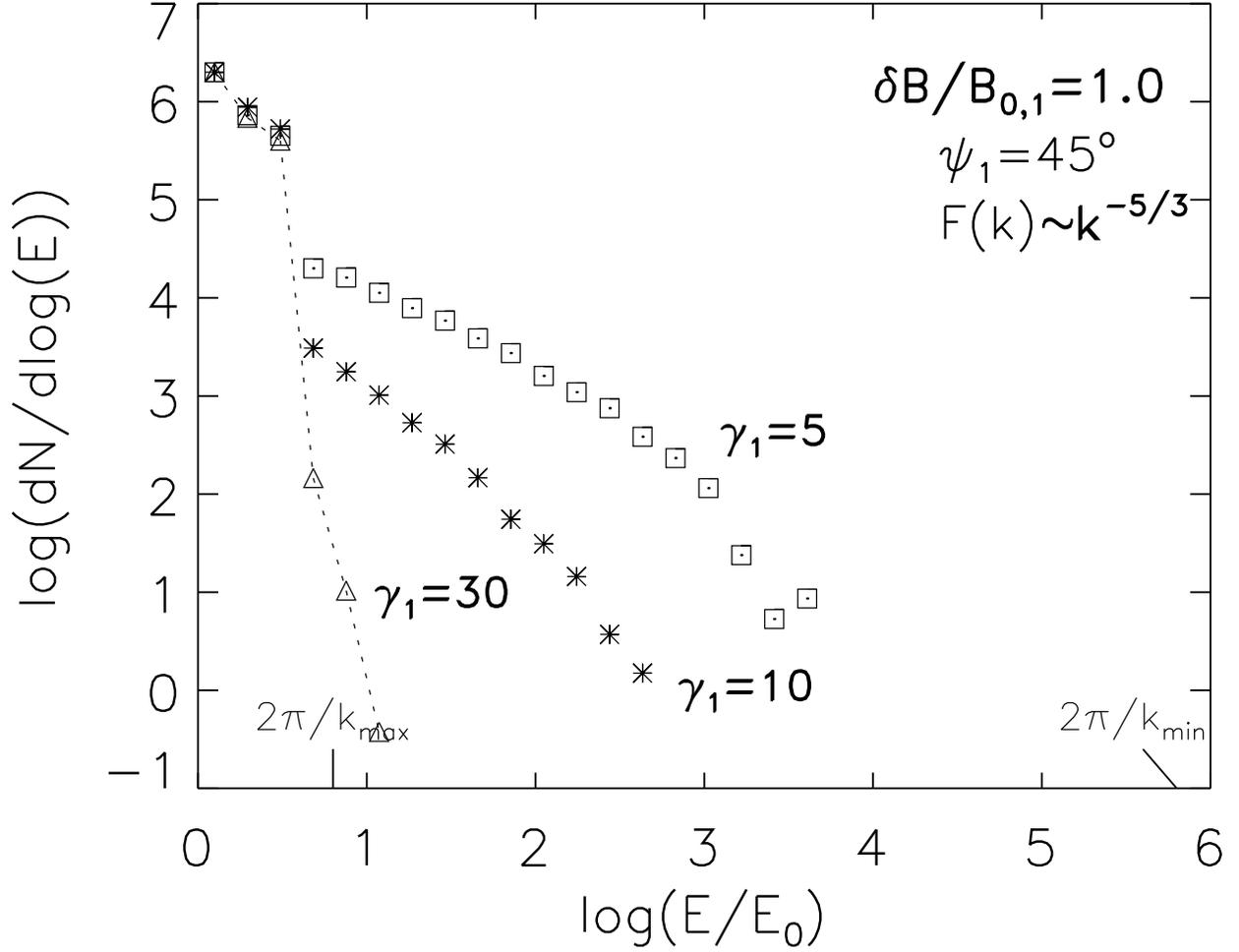}
\caption{Accelerated particle spectra at oblique superluminal shock waves for 
different shock Lorentz factors $\gamma_1$. The results are for the mean field 
inclination $\psi_1=45^o$ and the Kolmogorov wave power spectrum with 
$\delta B / B_{0,1} = 1.0$. 
\label{wobl2}}
\end{figure}

The characteristic features of particle acceleration processes at oblique
superluminal high-$\gamma$ shocks are illustrated in Figures \ref{wobl1} and 
\ref{wobl2}.\footnote{The results presented in Figure \ref{wobl1} for shocks 
with $\gamma_1=5$ correspond to those obtained with the previously used 
hybrid method for particle trajectory calculations 
\citep[see Fig. 9 in][]{nie04} and proves the validity of this approximate 
approach.} All injected particles are initially accelerated in a phase of 
``superadiabatic'' compression at the shock. Then, only a small fraction of 
these particles is further accelerated in the first-order Fermi process forming
an energetic tail in the energy spectrum. The tail shape and its  
extension to high energies strongly depend on the magnetic field turbulence 
spectrum. For the same power of magnetic field perturbations, the existence and 
extension of the tails depends on how much wave energy is contained in 
long-wave perturbations. For the flat wave power spectrum (with the spectral 
index $q = 1$; Fig. \ref{wobl1}{\it a}) the tails are steep and the resulting 
spectra of 
accelerated particles do not differ significantly from the pure compressed 
ones formed without any turbulence. For the high-amplitude 
Kolmogorov turbulence, with most power in long waves 
($q=5/3$; Fig. \ref{wobl1}{\it b}), much flatter high-energy tails are formed, 
which contain a substantial part of the accelerated particles energy density.  
These tails diverge from the power-law form by exhibiting a continuous 
steepening. For both types of wave spectra the cutoffs appear within the 
resonance energy range. To be noted from Figure \ref{wobl2} is that the cutoff 
energy decreases with growing shock Lorentz factor.

To understand these spectral features one has to consider the influence of 
the turbulence characteristics at energetic particle trajectories near 
the shock \citep[see, e.g.,][]{beg90,ost91,ost93,bed96,nie04}.
For the superluminal shock parameters considered here, with the projected velocity 
$u_{B,1} \equiv u_1/\cos{\psi_1} \approx 1.4c$ in each case, a low amplitude of 
magnetic field perturbations leads to the accelerated particle spectrum 
being just the anisotropically compressed upstream injected distribution. In our 
approach, with approximately monoenergetic particle injection, this process can
be seen in Figures \ref{wobl1} and \ref{wobl2} as a step-like spectral 
component at low energies. To enable the formation of a power-law tail at higher 
energies, some downstream particles must succeed in being transported back to the
shock for continued energization after the initial compression phase. There are 
two features of the perturbed magnetic field which can provide a backward 
transport. Either the resonant waves with a substantial amplitude are present, 
which enable particle cross-field diffusion backward to the mean plasma flow, or 
long-wave perturbations form locally subluminal field configurations near 
the shock and some particles propagating in these regions succeed in reaching the 
shock again. 

To better understand how these two factors can influence the acceleration process
let us analyze them separately. To enable efficient cross-field diffusion behind 
the shock, large-amplitude three-dimensional resonant MHD perturbations 
\citep[see, e.g.,][]{gia94,mich97,jon98} must be present.
In the case of relativistic shock waves, the shock moves in the downstream plasma
rest frame with velocity $u_2\approx c/3$ and the cross-field diffusion must 
approach its upper limit --- the Bohm diffusion efficiency --- in order to allow 
a sufficient fraction of downstream particles to remain continuously active in 
the energization process. This can be achieved in a relatively easy way by 
applying simplified turbulence modeling similar to those proposed in 
\citet{ost91}, \citet{ell90} or \citet{bed96,bed98}, where by selecting specific 
parameters for small-amplitude particle pitch-angle scattering one allows for 
conditions with nearly isotropic particle diffusion, or the modeling of 
\citet{ost93}, where upstream sinusoidal field perturbations are selected from 
the narrow, approximately resonance wavevector range at all particle energies. 
Studies involving finite wavevector range turbulence spectra, like the 
ones considered in \citet{nie04} and in the present paper, show a substantial
difficulty in generating the downstream field perturbations that allow
for efficient cross-field diffusion.

The presence of long-wave ($\lambda\gg r_g$) magnetic field perturbations of 
substantial amplitude changes the local shock conditions for particle 
acceleration. In superluminal shocks they form finite intermittent volumes near 
the shock with varying magnetic field obliquities, including the subluminal ones,
where the acceleration process can proceed more easily and in some cases particle 
reflections from the shock can occur (see \citet{nie04} for the discussion of these
effects in mildly relativistic shocks). However, for the in-average-perpendicular
field configuration such subluminal regions are advected with the plasma flow
and replaced by successive regions with possibly superluminal conditions, thus 
removing (advecting downstream) accelerated particles from the shock. 
Only the cross-field diffusion processes can redistribute some particles to 
new subluminal field configurations to extend the spectrum to higher energies. 
The injected particles that enter the acceleration process under superluminal
local conditions do not reach energies much higher than 
possible with the initial compression only. 

In the case of the flat wave power spectra (Fig. \ref{wobl1}{\it a}) and highly 
perturbed conditions ($\delta B/B_{0,1} = 1.0$ or 3.0 upstream of the shock) 
the amplitudes of resonant magnetic field perturbations are small upstream, 
and nonlinear downstream of the shock. However, the downstream compressed 
turbulence is strongly anisotropic, with most power in the magnetic field 
components parallel to the shock front,%
\footnote{This structure should relax to more isotropic 3D one further downstream,
but this relaxation process cannot be analyzed within the present approach.}
which results in decreased particle diffusion efficiency along the shock normal
(see Appendix). These downstream conditions are essentially independent of the 
initial particle energy, and the particle spectra for $E_i$ = $0.1$, 1 and 10, and
for $\delta B / B_{0,1} = 1.0$ in Fig. \ref{wobl1}{\it a} have the same shape 
with the compressed step-like parts followed by the cutoff tails. At each 
upstream--downstream--upstream cycle most of the particles finish the cycle 
being advected far downstream from the shock, and only a few privileged ones meet
suitable local magnetic field configurations that enable further energization. 
Thus the observed decrease in the cutoff tail inclination with particle energy, 
until a final sharp cutoff, results from this ``subjective'' selection of 
particles which can continue the acceleration and substitute the escaping ones
in the successive simulation cycles. Such a tail is 
absent in the weakly perturbed case ($\delta B/B_{0,1} = 0.3$), and the tails 
do not differ substantially between our mildly ($\delta B/B_{0,1} = 1.0$) and 
highly perturbed ($\delta B/B_{0,1} = 3.0$) cases. 

In cases involving the Kolmogorov wave power spectrum (Fig. \ref{wobl1}{\it b}),
the physical situation is substantially modified. Here  
the acceleration process after particle injection proceeds through the 
``superadiabatic'' shock compression phase, followed by the energetic spectral 
tail formation at the ensemble of shock conditions involving different local 
magnetic field inclinations, due to high amplitude of the long waves. The local 
regions with subluminal configurations are gradually advected downstream of the 
shock leading to steepening of the spectrum and a cutoff formation at 
$E_{cut} \sim  10^4 E_0$, independent of the particle injection energy, as 
exemplified by the spectra for $E_i$ = $0.1$, 1, 10 and 100, and
for $\delta B / B_{0,1} = 1.0$ in Fig. \ref{wobl1}{\it b}. The cutoff energy is 
much below the resonance energy for the longest upstream waves ($\sim 10^6 E_0$).
It is difficult to judge which changing background factor dominates in 
determining $E_{cut}$: the lack of large extended regions of subluminal 
conditions near the shock for high-energy particles, or the decreasing 
efficiency of diffusion along the local mean magnetic field due to compressed 
medium-amplitude waves perturbing the downstream particle trajectories. 

In  Figure \ref{wobl2}, one can observe a drastic reduction of the spectral tail 
and an accompanying decrease of $E_{cut}$ with increasing shock Lorentz factor. 
At large $\gamma_1$ the shock propagation velocity in the downstream plasma rest 
frame is essentially constant, $u_2 \approx c/3$. However, increasing $\gamma_1$ 
leads to a decreasing normal diffusion downstream of the shock,%
\footnote{For the cases shown in Fig. \ref{wobl2}, the shock compression factor 
$r = R\gamma_1/\gamma_2$ is 15.8, 29.5 and 86.4 for $\gamma_1=5$, $10$ and 
$30$, respectively.}
and that is the most important cause of decreasing of $E_{cut}$. An additional
factor responsible for the observed changes of $E_{cut}$ is a weaker particle
scattering upstream of the shock for growing $u_1 \rightarrow c$, due to the
shorter particle residence time before recrossing the shock downstream. 

\begin{figure}
\plotone{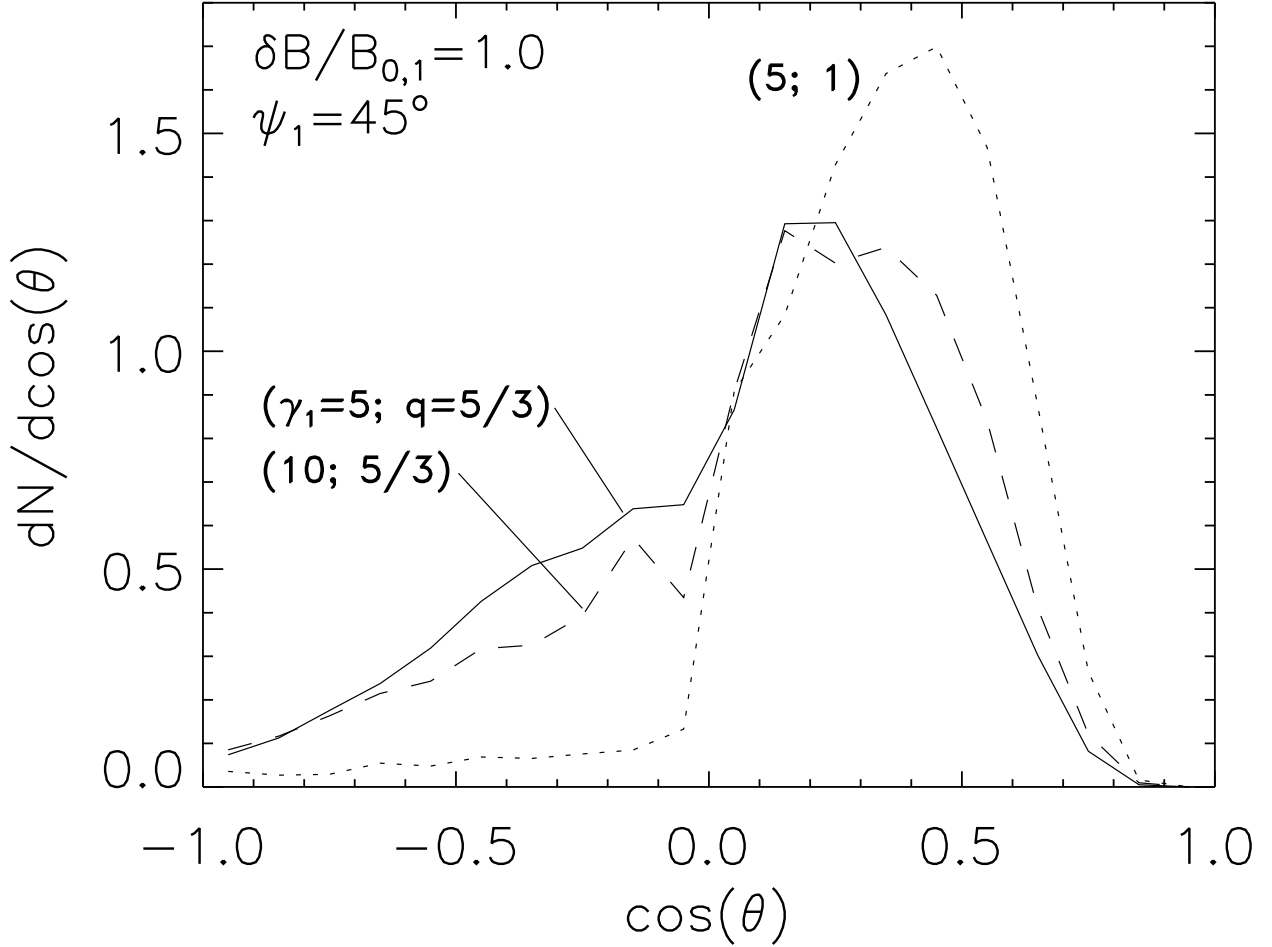}
\caption{Particle angular distributions at the superluminal shock waves
for $\delta B / B_{0,1}=1.0$ and $\psi_1=45^o$, in the shock rest frame.
The curves are calculated by summing the quantity $w/(|v_x|+0.005)$ in the
the respective $\cos\theta$ bins at every particle shock crossing, 
where $\theta$ is the angle between
the particle momentum and the shock normal, $v_x$ is a normal component of the 
particle velocity and $w$ is its weight. The low-energy particles, with larger
weights, thus provide the main contribution to the angular distributions.
Only particles forming energetic spectral tails in the energy spectra shown in 
Fig. \ref{wobl1} are included --- the distributions are averaged over particle 
energies and, therefore, approximate.
The angular distributions for the Kolmogorov wave power spectrum ($q=5/3$) and 
$\gamma_1=5$ and 10 are presented with the solid and dashed lines, respectively,
and for the flat wave power spectrum ($q=1$) and $\gamma_1=5$ by the dotted line.
All distributions are normalized to the unit surface area under the respective 
curves. Particles with $\cos\theta<0$ are directed upstream of the shock.
\label{angd1}}
\end{figure}

Figure \ref{angd1} shows particle angular distributions for the superluminal
shocks propagating in a mildly perturbed ($\delta B/B_{0,1} = 1.0$) magnetic field.
The distributions are calculated for particles forming energetic tails in the
spectra shown in Figures \ref{wobl1} and \ref{wobl2}. The differences in the
angular distributions for $\gamma_1=5$ and the Kolmogorov ({\em solid line}) and
the flat ({\em dotted line}) turbulence illustrate the variation of the 
physical conditions for particle acceleration discussed above for both types of 
the perturbed magnetic field power spectra. In particular, the effects of 
long-wave perturbations for $q=5/3$, which enable a larger fraction of particles 
to cross the shock back into the upstream region are apparent. 
The results for the Kolmogorov turbulence and for $\gamma_1=10$ 
({\em dashed line}) and $\gamma_1=5$ show the effects of the increasing shock
Lorentz factor on the angular distributions at the shock.

\subsection{Parallel High-$\gamma$ Shocks}

\begin{figure}
\plotone{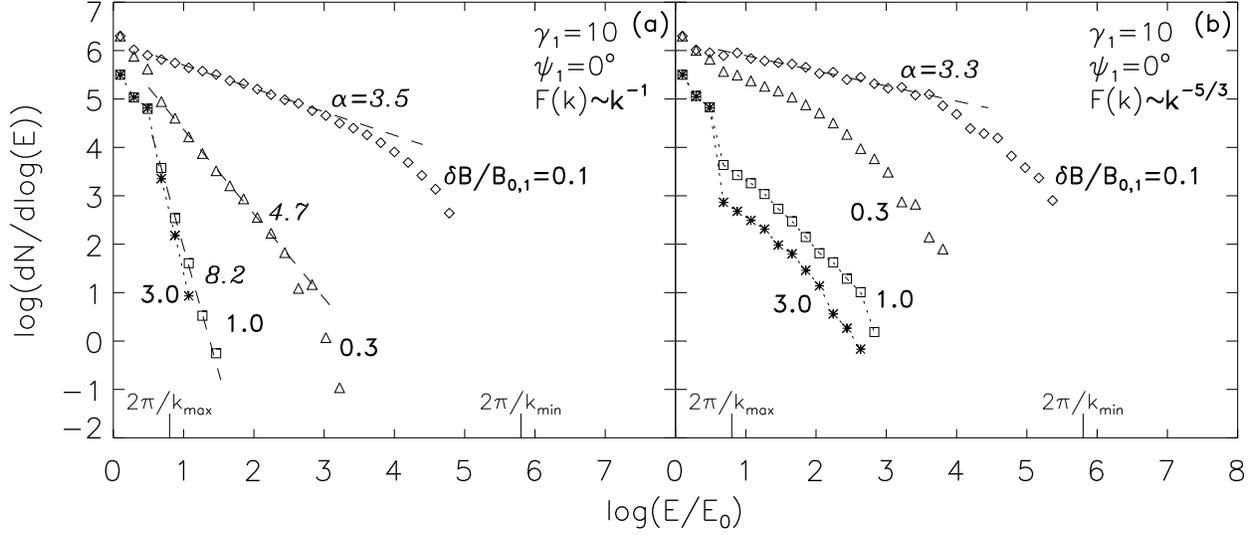}
\caption{Particle spectra at parallel shock waves with $\gamma_1 = 10$. 
The perturbed magnetic field parameters are given in each panel. Linear fits to 
the power-law parts of some spectra are presented, with the phase-space spectral 
indices values $\alpha$ given in italic.
\label{wpar1} }
\end{figure}

\begin{figure}
\plotone{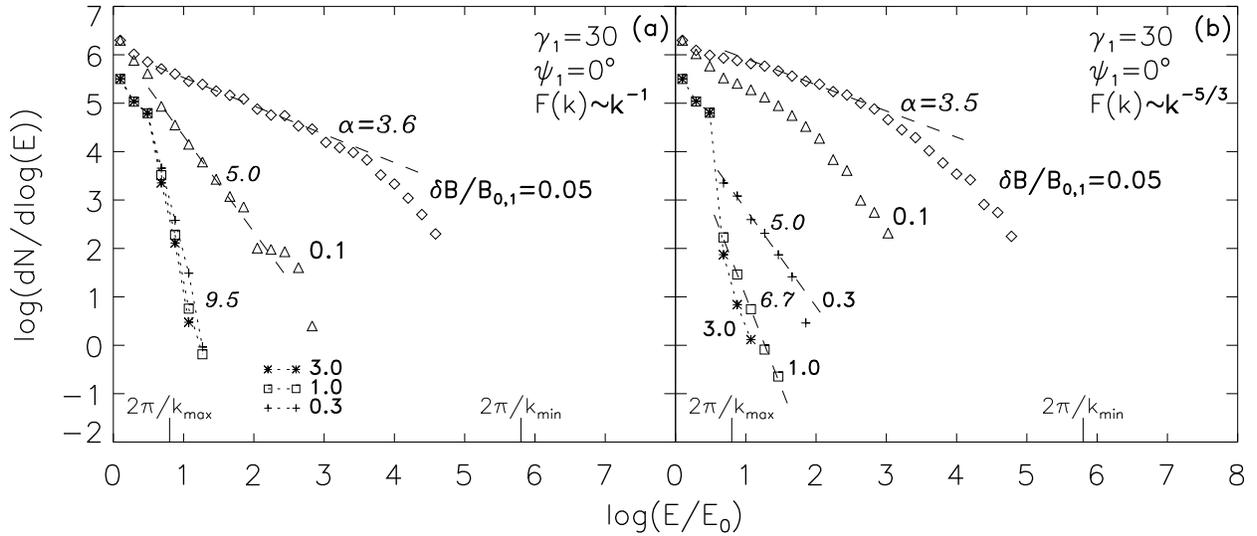}
\caption{Particle spectra at parallel shock waves with $\gamma_1 = 30$ (see
description of Fig. \ref{wpar1}). 
 \label{wpar2} }
\end{figure}

The effects of downstream magnetic field compression observed in particle spectra 
formed at oblique shocks are also found in parallel ultrarelativistic shocks, 
where the considered high-amplitude turbulence is subject to compression. 
Accelerated particle spectra for parallel shocks with $\gamma_1=10$ and 30 are 
shown in Figures \ref{wpar1} and \ref{wpar2}, respectively, for different 
amplitudes and power spectra of the magnetic field perturbations. 
For small perturbation amplitudes, very flat spectra ($\alpha < 4.0$), 
characteristic for oblique subluminal shocks, are generated. This is due to 
locally oblique magnetic field configurations formed by the long-wave 
perturbations at the shock front, and the respective field compressions 
downstream of the shock. The power-law parts of the spectra are 
followed by the cutoffs at high particle energies. The acceleration processes 
become less efficient for larger turbulence amplitudes. In these cases the 
acceleration 
mechanism resembles that acting at superluminal shocks: particle compression 
at the shock forms a step-like low-energy component of the spectrum, 
followed by the steeper (for the flat turbulence) or flatter (in the Kolmogorov 
case) spectral tails with cutoffs in the resonance energy range. By comparison 
of Figures \ref{wpar1} and \ref{wpar2} one notes that the critical turbulence 
amplitude allowing for efficient acceleration at parallel shocks is reduced
with increasing shock Lorentz factor. 

This character of the acceleration processes at ultrarelativistic parallel shocks
deviates from that predicted by numerical simulations using the small-angle 
scattering approximation for modeling of the perturbed magnetic field structure 
at the shock \citep{bed98,gal99,ach01,ell04} and confirmed by (semi)analytic 
approaches \citep{kir00,kes05}, as well as numerical models using other 
approaches \citep[e.g.,][]{lem03}. All these studies indicate formation of the
power-law particle spectrum with the asymptotic ($\gamma \rightarrow \infty$) 
spectral index $\alpha \approx 4.2 - 4.3$. However, as discussed in 
\citet{ostb02}, these acceleration models do not consider the effects of realistic 
turbulence at the shock, and in particular the significant role played by the 
long-wave perturbations. The different character of the acceleration processes 
at parallel shocks obtained in the present approach is an effect of the 
downstream magnetic field compression discussed above for oblique superluminal 
shocks. In the case of large-amplitude perturbations, the compression of 
tangential magnetic field components leads to the formation of an effectively 
perpendicular shock 
configuration. In such conditions, the efficiency of particle acceleration 
critically depends on the ability of particles to diffuse across field lines of 
the compressed turbulence. However, as the results for superluminal shocks 
show, the anisotropic diffusion involved is relatively inefficient for the 
compressed highly 
perturbed fields and the final particle spectra become very steep (see the 
results for $\delta B / B_{0,1} = 3.0$ and $1.0$ in Figure \ref{wpar1} and 
$\delta B / B_{0,1} = 3.0, 1.0$ and $0.3$ in Figure \ref{wpar2}). Note also, 
that features analogous to the ones observed in superluminal shocks --- 
dependence of the particle distribution on the shock Lorentz factor, variation 
of the spectral slope in the 
case of the Kolmogorov turbulence --- appear in the spectra for the parallel 
shocks. As in \citet{nie04}, particle spectra calculated for the considered 
small perturbation amplitudes are non-power-law ones in the full energy range, 
and the spectral index of the power-law part of the spectrum depends on the wave 
power spectrum of the magnetic field turbulence. The deviations of the spectra
from the power-law character are due to a limited dynamic range of the field
perturbations. Because of the lack of long-wave perturbations for high-energy
particles with $k_{res} < k_{min}$, locally oblique field configurations cannot
be formed and particles can be only transmitted downstream of the shock and 
escape.%
\footnote{Note that in calculations of the spectra for smallest perturbation
amplitudes we applied upper energy limits for the accelerated particles
(see \S 2.2; $E_{max,2}=3\cdot 10^3$ and $E_{max,2}=10^4$ for $\gamma_1=10$ and 
$\delta B/B_{0,1} =0.1$, and for the flat and Kolmogorov turbulence, respectively,
and  $E_{max,2}=2\cdot 10^3$ for $\gamma_1=30$ and $\delta B/B_{0,1} =0.05$),
because simulations of high-energy spectral tails require excessive run times.
Therefore, the extension of the spectral cutoffs to high energies was not 
reliably studied for these few cases.}

\begin{figure}
\plotone{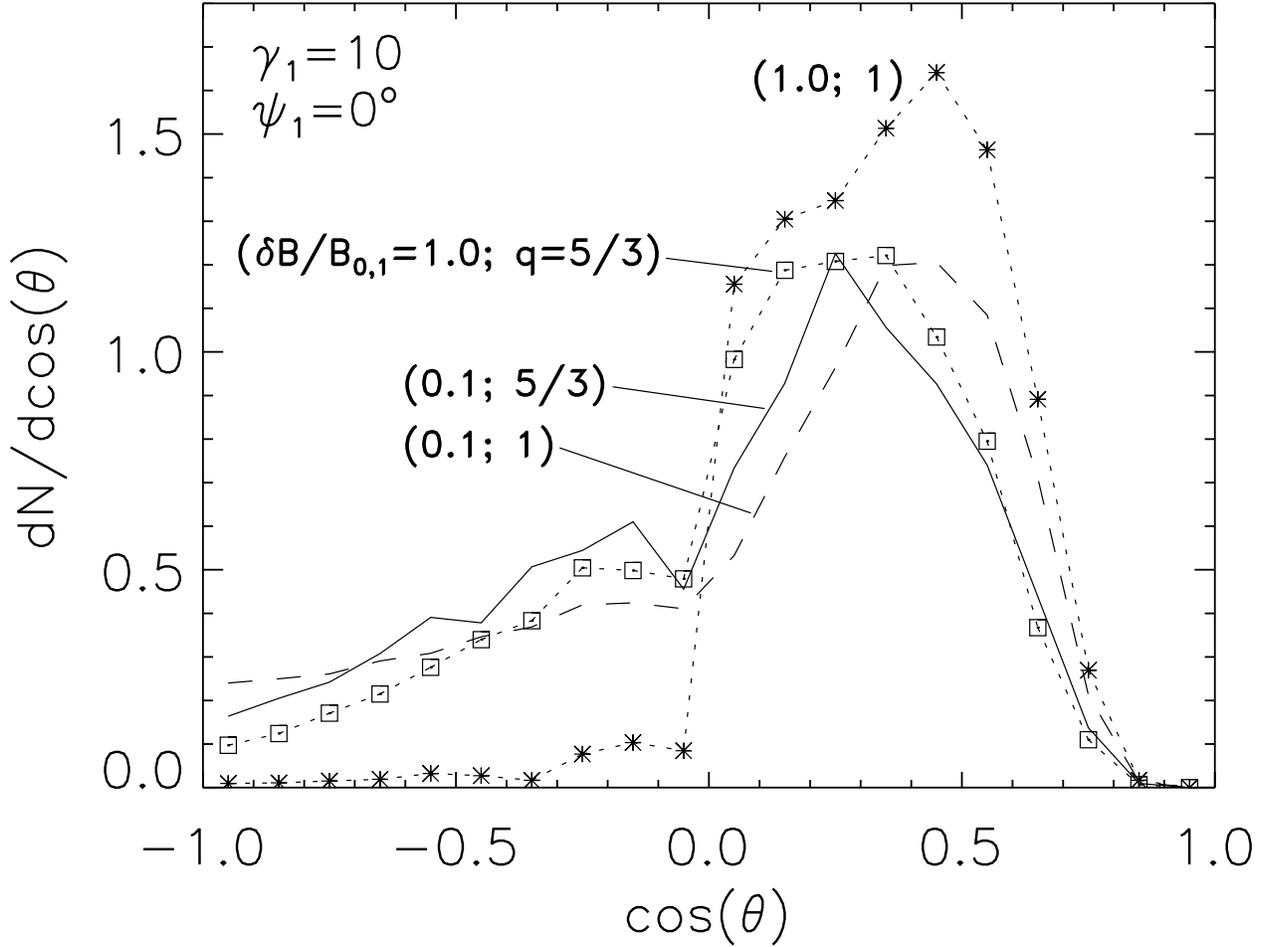}
\caption{Particle angular distributions at parallel shock waves
with $\gamma_1=10$. The distributions for the large turbulence amplitude
($\delta B / B_{0,1}=1.0$) are for particles forming energetic 
spectral tails (Fig. \ref{wpar1}) and are presented with squares and stars for
the Kolmogorov and the flat wave power spectrum, respectively. 
Angular distributions for the small-amplitude perturbations 
($\delta B / B_{0,1}=0.1$), shown with the solid ($q=5/3$) and dashed lines 
($q=1$), are formed by particles of energy $E_{s}\geq 1$ in the shock rest frame. 
\label{angd2}}
\end{figure}

The angular distributions for parallel shocks with $\gamma_1=10$ are presented
in Figure \ref{angd2}. For mildly perturbed ($\delta B/B_{0,1} = 1.0$) magnetic 
fields ($q=5/3$ -- {\em squares} and $q=1$ -- {\em stars}), one can see the 
features observed in the angular distributions for oblique superluminal shocks 
(Fig. \ref{angd1}). The differences in the distributions for small-amplitude
perturbations ($\delta B/B_{0,1} = 0.1$) reflect the difference in the spectral 
indices for power-law parts of the particle spectra for the flat 
({\em dashed line}) and Kolmogorov ({\em solid line}) turbulence. 

\section{SUMMARY AND DISCUSSION}

The results presented here for the first-order Fermi acceleration modeling at 
ultrarelativistic shock waves reveal several unknown earlier features of such 
processes. 
The most important finding of this work is that the generated particle spectra 
substantially depend on the form of the magnetic turbulence near the shock
--- contrary to the widely distributed opinion in the present literature. 
Simulations show a significant role of the perturbed magnetic field compression 
at the shock, which leads to highly anisotropic particle diffusion conditions in 
the approximately 2D turbulence formed downstream of the shock. In the case of the 
flat turbulence spectrum, the continuity of the magnetic field lines across the 
shock, and a limited particle diffusion along the shock normal in the downstream 
region result in the formation of very steep spectral tails with cutoffs within 
the resonance energy range considered. The spectra derived for the Kolmogorov 
turbulence show, that the existence of high-amplitude long-wave magnetic field 
perturbations can substantially modify the acceleration process, allowing for the
formation of more extended and flatter spectral components at superluminal shocks,
as well as very flat components at parallel shock waves. However, also in this 
case the spectrum steepening begins and/or the energy cutoff occurs a few orders 
of magnitude below the maximum particle energy corresponding to the 
resonance condition for the longest waves upstream of the shock. The simulations 
also show that for the same background conditions the shock waves with larger 
Lorentz factors produce steeper spectra with lower cutoff energies.

The compressed magnetic field turbulence structure downstream of the shock is 
expected to relax with time to the lower-energy isotropic 3D distribution. 
This may involve various plasma heating and cosmic ray acceleration processes, 
discussion of which is outside the scope of the present paper.
We note, however, that the characteristic time scale for such the isotropization 
at the given wavelength $\lambda$ can be estimated as $\tau_{iso} > \lambda/V_A$ 
($V_A < c$ is the Alfv\'en velocity in the downstream medium), so that the
relaxation processes cannot completely remove the considered anisotropy at the 
gyration time scale for particles with $r_g \la \lambda$. Thus, such particles 
are always influenced downstream of the shock by the respective resonance and 
long-wave {\em anisotropic} structure.

In the following paper (Niemiec \& Ostrowski, in preparation) we will present 
a further extension of the present model by including into consideration a 
highly nonlinear short-wave turbulent component, analogous the one formed at the 
shock due to plasma
instabilities discussed in the literature (see \S 2.1).

Our present results for ultrarelativistic shock waves, as well as the results 
obtained for mildly relativistic shocks \citep{nie04}, require revision of many 
earlier discussions of cosmic-ray acceleration up to very high energies 
at relativistic shock waves. The results show, that turbulent conditions near 
the shock which are consistent with the shock jump conditions can lead to 
substantial modifications of the acceleration picture with respect to the 
simplified models producing the wide-range power-law energy distributions, often 
with the ``asymptotic" or ``universal'' spectral index. With the turbulent 
magnetic field structures analyzed here, the generation of ultra--high-energy 
cosmic rays at ultrarelativistic shocks is ineffective. 
The role of the first-order Fermi process on the observational properties
of astrophysical sources hosting relativistic shock waves requires serious 
reanalysis. 

Recently we have become aware of the work by \citet{lem05}, who modified the
approach of \citet{lem03} to discuss particle acceleration at high-$\gamma$ 
shocks with the downstream magnetic field structure obtained from the shock jump 
conditions. Unfortunately, the numerical method used in both papers does not 
take into account the correlations between upstream and downstream 
trajectories at particle shock crossings, which are introduced by the
field structure. This is caused by the separate statistical 
averaging of the particle transport properties upstream and downstream, and the
subsequent treatment of these distributions as independent.
Therefore, the method cannot properly reproduce the resulting 
correlations in particle energy gains, return probabilities, 
and the respective propagation times at successive shock crossings  
\citep[e.g.,][]{bed96} and their effect on the final particle spectra.
The approach is thus unable, or only 
partially able, to reproduce the results obtained with more realistic 
background conditions discussed in \citet{nie04} and in the present paper.  

\acknowledgments
We are grateful to Martin Pohl for useful discussions of the results and for
providing valuable comments and suggestions on the manuscript. We also thank
David Carter-Lewis for carefully reading the manuscript.
The present work was supported by the Polish State Committee for Scientific 
Research in 2002-2005 as a research project PBZ-KBN-054/P03/2001, and
by MNiI in 2005-2008 as a research project 1 P03D 003 29.

\appendix
\section{PARTICLE DIFFUSION DOWNSTREAM OF THE SHOCK} 

In our simulations the magnetic field structure downstream of the shock becomes 
effectively two-dimensional, perpendicular to the shock normal. Due to the 
inefficient particle cross-field diffusion, advection of particles with the 
general downstream flow leads to high particle escape rates resulting in steep 
particle spectra. 

\begin{figure}
\plotone{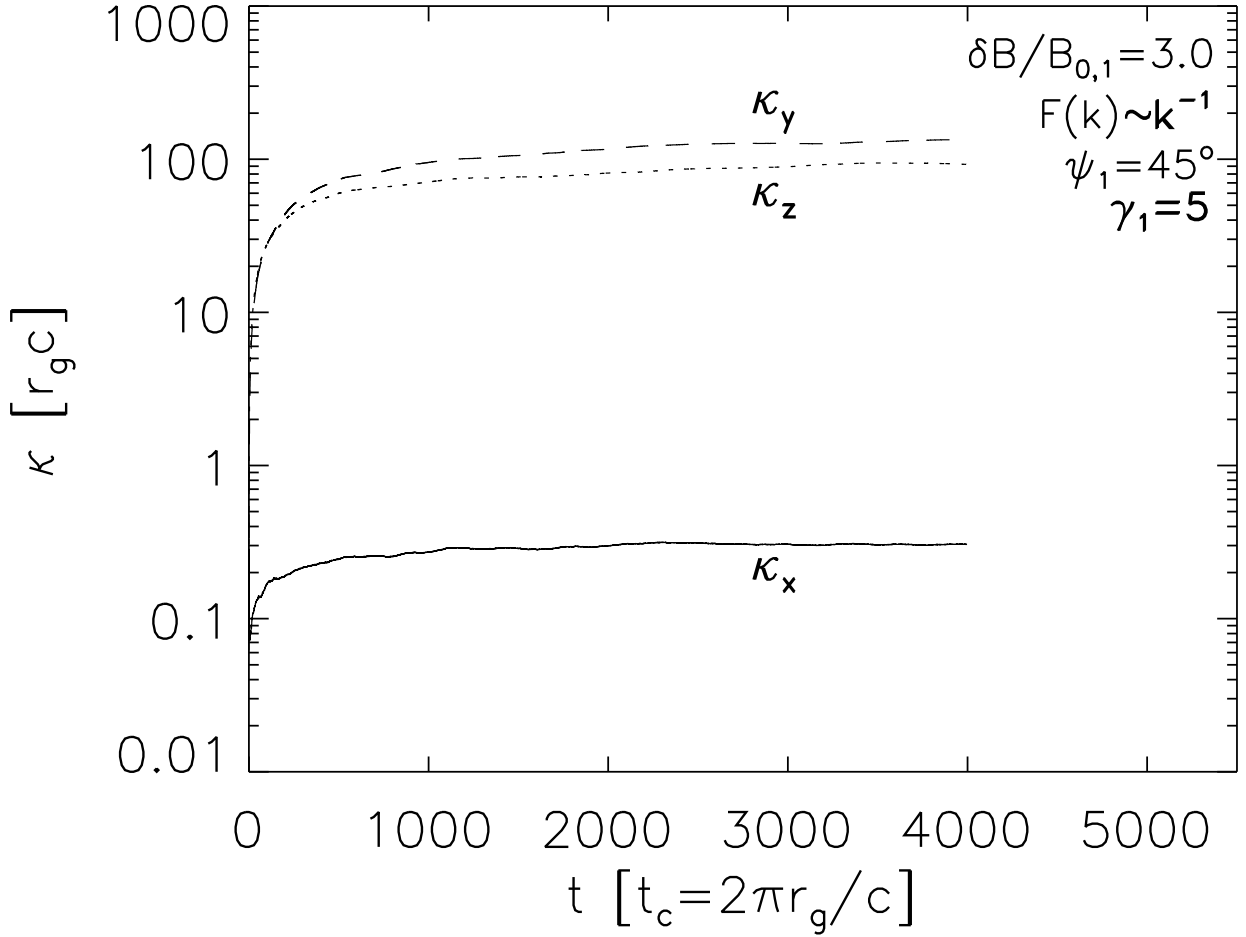}
\caption{Modeling of diffusion coefficients $\kappa_i \:(i=x,y,z)$ in the 
compressed perturbed magnetic field downstream of the shock for particles of 
energy $E_2=1$ and for the shock wave with $\gamma_1=5$; $t$ is the simulation time. 
The mean field inclination is $\psi_1=45^o$ and the
amplitude of the turbulent component with the flat power spectrum is 
$\delta B / B_{0,1}=3.0$ upstream of the shock. Particle gyroradius $r_g$, applied
as a measure of the diffusion coefficient and time unit, is calculated for the 
average downstream magnetic field. 
 \label{dyff} }
\end{figure}

To better understand propagation conditions in this region, we 
derived particle diffusion coefficients $\kappa_i \:(i=x,y,z)$ along the three 
axes in the downstream plasma rest frame, following the method described in 
\citet{ost93} \citep[see also][]{mich97}. 
The modeling shows a highly anisotropic diffusion downstream of
the shock, with the coefficient along the shock normal 
$\kappa_x \ll \kappa_y$, $\kappa_z$. For the example considered in Fig. \ref{dyff}
of the shock wave with $\gamma_1=5$, propagating in the magnetic field with
the mean field inclination $\psi_1=45^o$ and the flat wave power spectrum of the 
(upstream) amplitude $\delta B / B_{0,1} = 3.0$, the downstream diffusion 
coefficients for particles of energy $E=1$ are 
$\kappa_y / \kappa_x \sim 4 \cdot 10^2$ and 
$\kappa_z / \kappa_x \sim 3 \cdot 10^2$. The difference between $\kappa_y$ and 
$\kappa_z$ comes from the contribution of the mean field component which lies in 
the $x-y$ plane ($\psi_2=86^o$ in this case). The exact values of the
diffusion coefficients depend on particle energy.
One should note, however, that the diffusion coefficient values derived here can
serve only as a general characterization of the propagation properties downstream
of the shock. They represent particle motion on time scales much larger than
the particle gyration time scale, the latter being important in the modeling of 
the relativistic shock acceleration.

\clearpage

\end{document}